# LOW-TEMPERATURE FERROMAGNETISM IN A WEAKLY DOPED HUBBARD MAGNET

E.E. ZUBOV

O.O. Galkin Institute for Physics and Engineering
Nat. Acad. of Sci. of Ukraine

(72, R. Luxemburg Str., Donetsk 83114, Ukraine; e-mail: zubov@kinetic.ac.donetsk.ua)

©2010

In the framework of the diagrammatic method with self-consistent field, the maximum on the temperature dependence of the susceptibility of a weakly doped narrow-band Hubbard magnet below the Curie temperature $T_C$ is predicted. By numerical calculations, it is proved that it appears at small hole concentrations $n_h$. In this case, the temperature dependence of the magnetization $M(T)$ has a typical Fermi-like shape with the point of inflection at a temperature $T_{\rm inf}$. The approximate solution of the system of equations for the mean spin and the chemical potential gives the Schottky susceptibility typical of a two-level system with a gap of order of $n_h(1-n_h)W$, where W is the bandwidth. This behavior reflects the existence of two subbands with up- and down-spins. It may be observed experimentally in the surface layers of oxide metal nanoparticles with narrow bands and a weak oxygen nonstoichiometry.

At present, the important role of nanoparticle surfaces in the magnetic ordering of the electron subsystem with a narrow band is well established. In particular, a model based on the ideal inner core and an outer shell of nanoparticles with oxygen nonstoichiometry, vacancies, and stresses was suggested for ferromagnetic metallic compound $La_{2/3}Ca_{1/3}MnO_3$ in work [1]. A weak surface ferromagnetism is a universal peculiarity of many oxide nanoparticles including nonmagnetic $Al_2O_3$, $ZnO$, and $CeO_2$ [2]. A complicated character of the interaction between surface and volume magnetic ions in some cases causes the exchange bias in nanostructures [3].

The increase of the susceptibility $\chi(T)$ at lowering the temperature $T$, a sharp drop of $\chi(T)$ with increase in a magnetic field, and the maximum in the curve $\chi(T)$ at a definite doping parameter $x$ and a temperature $T = T_{\max}$ were observed earlier in HTSC compounds $RBa_2Cu_3O_{6+d}$ (where R = Y, Nd, Sm) in the dielectric phase [4–7]. In work [8], the experimental results were interpreted, by considering the competition of the in-plane antiferromagnetic (AFM) and the weak ferromagnetic (FM) interplanar exchange, whose origin is related to the interaction between the quadrupole center and virtual magnons of the AFM matrix. The maximum of $\chi(T)$ was established [9] to occur in strongly correlated electron systems as a result of correlations between localized spins in the AFM phase. With increase in $x$, the maximum shifts to the area of lower temperatures, which points out the strengthening of a destructive influence of the hole dynamics on the AFM order.

To explain the phenomena observed in $RBa_2Cu_3O_{6+d}$ compounds, we will suppose the existence of a heterogeneous magnetic structure in grains of the ceramics under study. This magnetic heterogeneity is seemingly related to the structural inhomogeneity caused by the influence of a grain boundary. As a result, the oxygen index can vary at moving from the grain center to the boundary. Apparently, this gives the excess of the hole content on the surface. Therefore, in spite of the fact that the rest volume has AFM order, it is destroyed near the grain surface with forming a weak FM or paramagnetic state (PM). The realization of a FM or PM state depends on the degree of doping and the relation between the kinetic energy of electrons and their indirect exchange interaction [10–12]. The comparison of calculated and experimental values of $\chi(T)$ will allow us to make a conclusion about the surface ferromagnetism in $RBa_2Cu_3O_{6+d}$.

For the description of the low-temperature thermodynamics of electrons with a narrow band, the theory of effective field [11–13] was proposed. By means of the scattering matrix formalism in the first order of perturbation theory with the formal parameter $t$ (hopping integral), one can find correctly the kinematic contribution to the magnetization and the susceptibility of a weakly doped Hubbard magnet. The system of equations describing the mean spin $\langle S^z \rangle$ and the chemical potential $\mu$ as functions of $T$, the bandwidth $W$, and the electron concentration $n$ is as follows:

$$\langle F^{\sigma 0}\rangle = \langle F^{\sigma 0}\rangle_1 - \frac{1}{N}\sum_q f(E_{q\sigma}^F) + f(\varepsilon_\sigma). \quad (1)$$

Here, the mean combined population $\langle F^{\sigma 0}\rangle = 1 - \frac{n}{2} + \sigma\langle S^z\rangle$, $\sigma=\pm 1$, $h$ is a magnetic field, $\varepsilon_\sigma = -\mu - \frac{1}{2}\sigma h -$

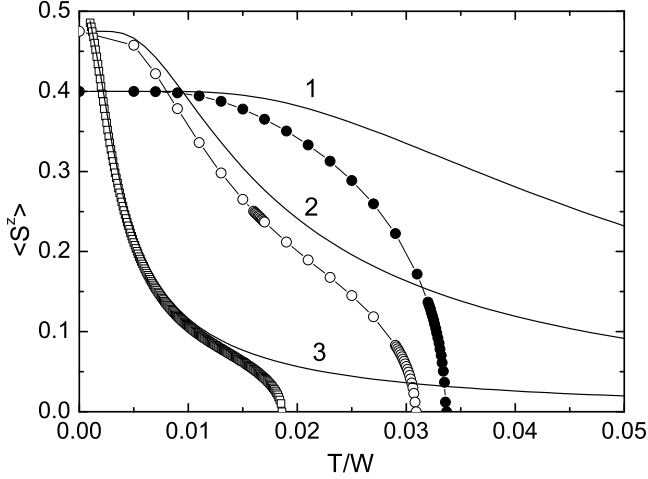

Fig. 1. Temperature dependences of the mean electron spin at $\kappa=0$ and $n = 0.8, 0.95$ and $0.99$ (solid curves 1-3 correspond to Eq. (3)). The dotted curves were obtained from Eqs. (1) by numerical calculations

$\kappa t(0)\left\{\frac{n}{2} - \sigma \langle S^z \rangle\right\}$, $\kappa$ is the exchange parameter, $f$ is the Fermi distribution function, and

$$E^F_{q\sigma} = \varepsilon_\sigma + t(q)\langle F^{\sigma 0}.\rangle$$

Here, $t(q)$ is the Fourier transform of the hopping integral. The "dressed" population $\langle F^{\sigma 0}\rangle_1$ can be expressed as

$$\langle F^{\sigma 0}\rangle_1 = \frac{e^{\beta E_\sigma} + 1}{1 + e^{\beta E_\sigma} + e^{\beta E_{-\sigma}}}, \qquad (2)$$

where $E_\sigma = -\varepsilon_\sigma + \delta\mu^F_{-\sigma}$ and $\delta\mu^F_\sigma == \frac{1}{N}\sum_q \beta t(q) f(E^F_{q\sigma})$.

In the limit of low temperatures $T \ll T_C$, where $T_C$ is the Curie temperature, the system of equations (1) was solved analytically [12]. In particular, we obtain

$$\langle S^z \rangle = \frac{n}{2} - f(\tilde{\varepsilon}), \qquad (3)$$

where $\tilde{\varepsilon} = Wn(1-n)/2 + h - W\kappa\langle S^z\rangle.$ It follows from Eq. (3) that a deviation $\langle S^z \rangle$ from the saturated magnetization obeys the Fermi distribution with the parameter $\tilde{\varepsilon}$ characterizing a splitting of the bands with up- and down-spins. This splitting is formed by the applied magnetic and effective kinematic fields competing with a self-consistent field of the exchange origin.

At $h=0$ and $\kappa = 0$, the parameter $\tilde{\varepsilon}$ depends on the factor $n(1-n)$, whose value may be small at $n \ll 1$. In this case, the thermal fluctuations can have a pronounced effect on the Nagaoka's state. In particular, the temperature dependence of the magnetization changes from the typical Fermi-like behavior with the inflection point at $T_{\max}$ to the Brillouin-type behavior without it. From (3), it is easy to find

$$\chi(T) = \frac{\partial \langle S^z \rangle}{\partial h} = \frac{1}{4T}\cosh^{-2}\left[\frac{Wn(1-n) + 2h}{4T}\right], \qquad (4)$$

where $\kappa = 0$. Susceptibility (4) has a Schottky-like maximum at $T = T_{\max}$ determined from the equation

$$T_{\max}/W = 0.5 n(1-n) \tanh\left(\frac{Wn(1-n)}{4T_{\max}}\right). \qquad (5)$$

Unfortunately, the approximate solution (3) contains the inflection point always because it works at low temperatures. The exact numerical solution of Eqs. (1) at various concentrations $n$ shows a change of the shape of $\langle S^z \rangle$ as a function of $T$ at the definite concentration $n$ (Fig. 1). To understand the reason for such a modification, we will obtain analytical expressions for the longitudinal molar susceptibility $\chi_\parallel(T)$ from (1). The differentiation of (1) with respect to $T$ gives

$$\chi_\parallel(T) = \frac{\mu_B^2 g^2 N_A}{k_B}\frac{b_\sigma a_{-\sigma\sigma} - b_{-\sigma} a_{\sigma-\sigma}}{a_{\sigma\sigma}a_{-\sigma\sigma} - a_{-\sigma-\sigma}a_{\sigma-\sigma}}, \qquad (6)$$

where $\mu_B$ is the Bohr magneton, $g$ is the $g$-factor along the $z$ axis, $N_A$ is the Avogadro constant, and $k_B$ is Boltzmann's constant. We also have

$$\begin{aligned}
a_{\sigma\sigma} &= \sigma\left(1 - \beta^2\psi_{-\sigma}F_{\sigma-\sigma} - \beta^2\psi_\sigma F_\sigma + \beta\varphi_{-\sigma}\right) \\
a_{\sigma-\sigma} &= \beta(F_\sigma(1+\beta\varphi_\sigma) - F_{\sigma-\sigma}(1+\beta\varphi_{-\sigma}) + \\
&\quad + \Delta_{-\sigma} - f(\varepsilon_{-\sigma})(1 - f(\varepsilon_{-\sigma}))), \\
b_\sigma &= \sigma\beta(F_{\sigma-\sigma}(1 - \beta\varphi_{-\sigma}) + F_\sigma(1 - \beta\varphi_\sigma) + \\
&\quad + \Delta_{-\sigma} - f(\varepsilon_{-\sigma})(1 - f(\varepsilon_{-\sigma}))).
\end{aligned} \qquad (7)$$

Here,

$$\begin{aligned}
\varphi_\sigma &= \frac{1}{N}\sum_q t(q) f(E^F_{q\sigma})\left(1 - f(E^F_{q\sigma})\right), \\
\psi_\sigma &= \frac{1}{N}\sum_q t^2(q) f(E^F_{q\sigma})\left(1 - f(E^F_{q\sigma}))
\end{aligned} \qquad (8)$$

and

$$\Delta_\sigma = \frac{1}{N}\sum_q f(E^F_{q\sigma})\left(1 - f(E^F_{q\sigma})\right), \qquad (9)$$

$$F_\sigma = \frac{e^{\beta E_{-\sigma}}\left(e^{\beta E_\sigma} + 1\right)}{\left(1 + e^{\beta E_\sigma} + e^{\beta E_{-\sigma}}\right)^2},$$

$$F_{\sigma-\sigma} = F_{-\sigma\sigma} = \frac{e^{\beta E_{-\sigma}} e^{\beta E_\sigma}}{\left(1 + e^{\beta E_\sigma} + e^{\beta E_{-\sigma}}\right)^2}, \qquad (10)$$

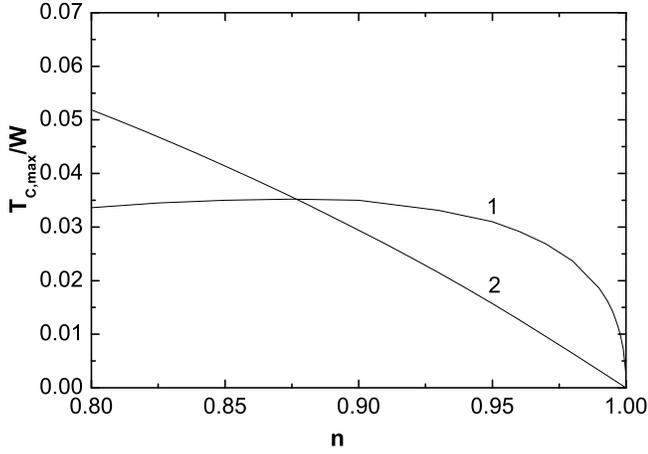

Fig. 2. Concentration dependences of $T_C$ (curve *1*) and $T_{\max}$ (curve *2*)

where $\beta = 1/T$.

In the PM phase $\langle S^z \rangle = 0$, and functions (8)–(10) do not depend on the index $\sigma$: $\varphi_\sigma = \varphi$, $\psi_\sigma = \psi$, $\Delta_\sigma = \Delta$, $\epsilon_\sigma = \epsilon$, $E_\sigma = E$, $F_\sigma = F$, $F_{\sigma-\sigma} = F_{\uparrow\downarrow}$. Therefore, Eq. (6) is simplified to

$$\chi_\parallel(T) = \frac{\mu_B^2 g^2 N_A}{k_B} \times$$

$$\times \frac{(F + F_{\uparrow\downarrow})(T - \varphi) + T\Delta - Tf(\varepsilon)(1 - f(\varepsilon))}{T^2 - \psi(F + F_{\uparrow\downarrow}) + T\varphi}. \quad (11)$$

The equality of the denominator in (11) to zero yields the following expression for the Curie temperature:

$$T_C^2 = \frac{1}{2 + e^{-\frac{\mu + \delta\mu_0}{T_C}}} \frac{1}{N} \sum_q t^2(q) f(E_q) [1 - f(E_q)] -$$

$$-\frac{1}{N} \sum_q t(q) f(E_q)[1 - f(E_q)]. \quad (12)$$

Here, $E_q = -\mu + \left(1 - \frac{n}{2}\right) t(q)$ and $\delta\mu_0 = \frac{1}{N} \sum_q t(q) f(E_q)$. Setting $\langle S^z \rangle = 0$ in (1), we can obtain the chemical potential $\mu$ as a solution of the equation

$$1 - n = \frac{1}{1 + 2e^{\frac{\mu + \delta\mu_0}{T}}} - \frac{2}{N} \sum_q f(E_q) + 2f(-\mu).$$

Expanding the denominator of $\chi_\parallel(T)$ in a series in $T - T_C$, one can find the Curie–Weiss law

$$\chi_\parallel(T) = \frac{C}{T - T_C}, \quad (13)$$

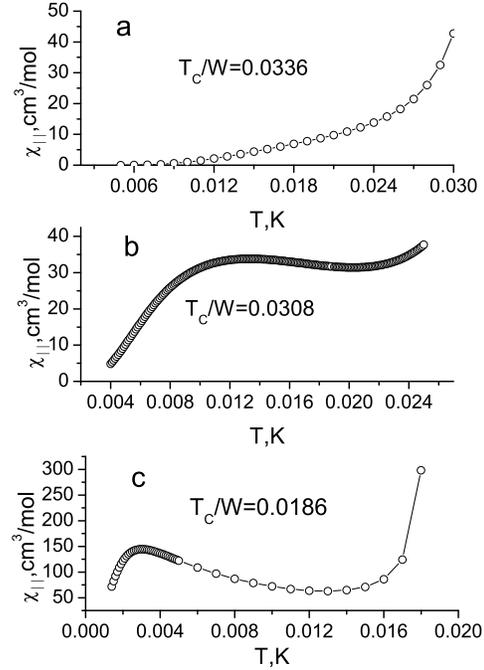

Fig. 3. Temperature dependences of the molar susceptibility $\chi_\parallel(T)$ at $\kappa=0$ and concentrations $n = 0.8$ (*a*), 0.95 (*b*), and 0.99 (*c*)

where the Curie constant has the form

$$C = \frac{\mu_B^2 g^2 N_A}{k_B} \times$$

$$\times \frac{(F + F_{\uparrow\downarrow})(T_C - \varphi) + T_C \Delta - T_C f(-\mu)(1 - f(-\mu))}{Q(T_C)}. \quad (14)$$

In Fig. 1, the comparison of the values of mean spin calculated from the approximate equation (3) (lines) and the exact equation (1) (points) was made. It follows from Fig. 1 that the character of the temperature dependence of $\langle S^z \rangle$ is changed with increase in the electron concentration.

Using Eq. (12), one can find the concentration dependence of $T_C$ presented in Fig. 2. Here, the function $T_{\max}(n)$ obtained by solving Eq. (5) is displayed. From Fig. 2, it is seen that $T_{\max} > T_C$ at $n<0.88$. Therefore, the Schottky-like anomaly in $\chi_\parallel(T)$ disappears at these concentrations. This claim is supported by the results of numerical calculations of (11).

Figure 3 shows that, at $n=0.8$, the maximum on curve $\chi_\parallel(T)$ is absent. Moreover, the absolute values of sus-ceptibility (11) exceed essentially the experimentally observed ones for $SmBa_2Cu_3O_{6+d}$ [7]. This indicates that the volume, in which a weak ferromagnetism is observed, is small.

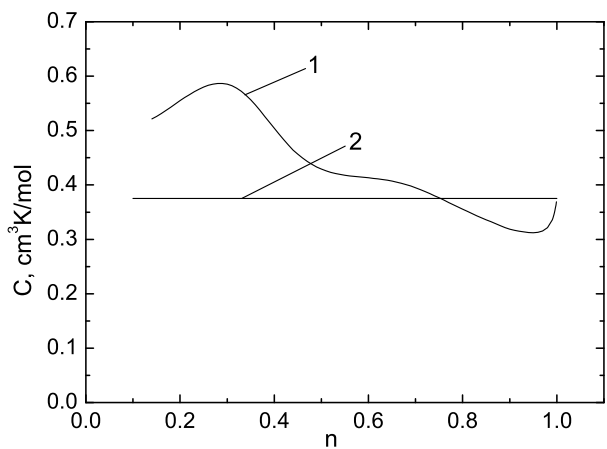

Fig. 4. Concentration dependence of the Curie constant $C$ (curve *1*). Straight line *2* corresponds to $\frac{1}{4}\frac{\mu_B^2 g^2 N_A}{k_B}$ for localized electron spins

In Fig. 4, the concentration dependence of the Curie constant is presented. One can see that, at $n=1$, we have the evident result $C = \frac{1}{4}\frac{\mu_B^2 g^2 N_A}{k_B}$ corresponding to localized electron spins, and $C$ as a function of $n$ has non-monotone character. The mentioned dependence can be used to find the electron concentration knowing $C$ from experiments.

Hence, the temperature dependence of the longitudinal susceptibility of a Hubbard magnet at low temperatures in the limit of a weak doping shows the Schottky-like anomaly. We suppose that the condition for the appearance of this anomaly is related to a small value of the mean thermal energy of electron excitation from the down- to spin-up subband. If this energy is comparable with the mean electron kinetic energy of FM ordering, the indicated anomaly disappears.

НИЗЬКОТЕМПЕРАТУРНИЙ ФЕРОМАГНЕТИЗМ У СЛАБКО ЛЕГОВАНОМУ ХАББАРДОВСЬКОМУ МАГНЕТИКУ

*Є.Є. Зубов*

Р е з ю м е

У межах теорії ефективного самоузгодженого поля для слабко легованого хаббардовського магнетика передбачено появу низькотемпературного максимуму типу Шотткі на кривій сприйнятливості. Показано, що дану аномалію зумовлено конкуренцією кінематичних і теплових взаємодій колективізованих електронів. Оцінка величини молярної сприйнятливості дозволяє зробити висновок про можливу реалізацію зазначеного явища в поверхневих шарах металооксидних наночастинок із вузькими зонами.